\documentstyle[aps,twocolumn,psfig,floats]{revtex}  
\def\beqa{\begin{eqnarray}}
\def\eqar{\end{array}}
\def\beqar{\begin{array}}
\def\eqa{\end{eqnarray}}
\def\bars{\begin{eqnarray*}}
\def\ears{\end{eqnarray*}}

\def \sm {Standard Model }
\def \susy {supersymmetry }
\def \susyq {supersymmetric }

\def \sugra {supergravity }
\def \RPV {R-parity violating }
\newcommand{\rpv}{\mbox{$\not \hspace{-0.15cm} R_p$}}

\def\l {\lambda }

\def \g {\gamma }

\def \bea {\begin{equation} }
\def \eea {\end{equation} }

\def \Eslash {E \kern-.5em\slash }
\def \pslash {p \kern-.5em\slash }
\def \kslash {k \kern-.5em\slash }

\newcommand{\bt}{\begin{tabular}}
\newcommand{\et}{\end{tabular}}
\newcommand{\bd}{\begin{displaymath}}
\newcommand{\ed}{\end{displaymath}\noindent}
\newcommand{\ec}{\end{center}}
\newcommand{\bc}{\begin{center}}

\draft

\begin{document}
\twocolumn[\hsize\textwidth\columnwidth\hsize\csname@twocolumnfalse\endcsname
\title{Resonant sneutrino production at Tevatron Run II}
\author{G. Moreau, M. Chemtob} 
\address{Service de Physique Theorique, CEA/Saclay,
F-91191 Gif-sur-Yvette Cedex, France}
\author{F. D\'eliot, C. Royon, E. Perez} 
\address{Service de Physique des Particules, DAPNIA, CEA/Saclay, 
91191 Gif-sur-Yvette Cedex, France}
\date{\today}

\maketitle


\begin{abstract}
We consider the single chargino production at Tevatron 
$p \bar p \to \tilde \nu_i \to \tilde \chi^{\pm}_1 l_i^{\mp}$
as induced by the resonant sneutrino production via a dominant 
\RPV coupling of type $\l'_{ijk} L_i Q_j D_k^c$.
Within a supergravity model, we study the three leptons final state.
The comparison with the expected background 
demonstrate that this signature allows to
extend the sensitivity on the \susyq mass spectrum 
beyond the present LEP limits and to probe 
the relevant \RPV coupling down to values one order of magnitude smaller 
than the most stringent low energy indirect bounds. The trilepton signal 
offers also the opportunity to reconstruct the neutralino mass in a 
model independent way with good accuracy.    
\end{abstract}

]




In the minimal supersymmetric standard model (MSSM), 
the \susyq (SUSY) particles must be produced in pairs.
In contrast, the single superpartner 
production which benefits from a
larger phase space is allowed in the \RPV (\rpv)
extension of the MSSM.
In particular the SUSY particle resonant production 
can reach high cross-sections
either at leptonic \cite{Han} or hadronic colliders \cite{Dim2}, 
even taking into account the strongest 
low-energy bounds on \rpv coupling constants \cite{Drein}.
Hadronic colliders provide an additional advantage in that they
allow to probe a wide mass range of the new resonant particle,
due to the continuous energy distribution of the colliding partons.
Furthermore, since the resonant production has a cross-section which is 
proportional to the relevant coupling squared,
this could allow an easier
determination of the \rpv coupling than pair production reaction.
Indeed in the latter case,
the sensitivity on the \rpv coupling is
mainly provided by the displaced vertex analysis for the 
Lightest Supersymmetric Particle (LSP) decay,
which is difficult experimentally especially at hadronic colliders.

The SUSY particle produced at the resonance 
mainly decays through 
R-parity conserving interactions into the LSP,
via cascade decays. 
In the case of a dominant 
$\l''_{ijk} U_i^c D_j^c D_k^c$ coupling, the decay of the LSP 
leads to multi-jets final states, wich have an high QCD background
at hadronic colliders. Besides, at hadronic colliders, the 
$\l_{ijk} L_i L_j E_k^c$ 
couplings do not contribute to resonant production.
In this letter, we thus assume a dominant 
$\l'_{ijk} L_i Q_j D_k^c$ coupling which initiates 
the resonant sneutrino production $\bar d_j d_k \to \tilde \nu_i$
and hence the single chargino production at Tevatron through
$ p \bar p \to \tilde \nu_i \to \tilde \chi^{\pm}_1 l_i^{\mp}$.
We focus on the three leptons signature 
associated with the cascade decay 
$\tilde \chi^{\pm}_1 \to \tilde \chi^0_1 l^{\pm}_p \nu_p$, 
$\tilde \chi^0_1 \to l^+_i \bar u_j d_k \ +c.c.$,
assuming the $\tilde \chi^0_1$ to be the LSP. 
The main motivation rests on the possibility to reduce the background.
This is similar in spirit to a recent study \cite{Rich} 
of the like sign dilepton signature from the single neutralino 
production at Tevatron via the resonant charged slepton production.


\begin{figure}
\begin{center}
\leavevmode
\centerline{\psfig{figure=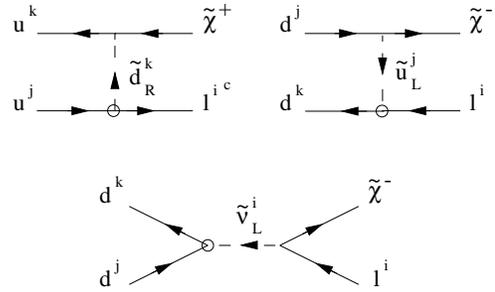,height=1.5in}}
\end{center}
\caption{Feynman diagrams for the single chargino 
production at Tevatron via the $\l'_{ijk}$ coupling 
(symbolised by a circle in the figure). 
The arrows denote flow of the particle momentum.}
\label{fig0}
\end{figure}

We concentrate on the $\l'_{211}$ coupling.
The associated hard scattering processes,
$ d \bar d \to \tilde \nu_{\mu} \to \tilde \chi^{\pm}_1 \mu^{\mp}$,
$d \bar d \to \tilde \chi^{\pm}_1 \mu^{\mp}$ and
$u \bar u \to \tilde \chi^{\pm}_1 \mu^{\mp}$ (see Fig.\ref{fig0}),
involve first generation quarks for the initial partons. 
The indirect constraint on 
this coupling is $\l'_{211}<0.09 (\tilde m /100GeV)$  \cite{Drein}.
While $\l'_{111}$ is disfavored due to severe constraints 
\cite{Drein}, 
the case of a dominant $\l'_{311}$ could also be of interest.

Our framework is the so-called minimal \sugra model (mSUGRA),
in which the absolute value of the 
Higgsino mixing parameter $\vert \mu \vert$ is determined by the 
radiative electroweak symmetry breaking condition.
We restrict to the infrared fixed point region for the top quark 
Yukawa coupling, in which $\tan \beta$ is fixed \cite{Pok}.
We shall present results for the low solution $\tan \beta \simeq 1.5$  
and for $sign(\mu)=-1$, $A=0$. 
In fact the cross-section for the single chargino production
depends smoothly on the $\mu$, $A$ and $\tan \beta$ parameters.
The cross-section can reach values of order a few picobarns. 
For instance, choosing the mSUGRA point, $M_2(m_Z)=200GeV$, 
$m_0=200GeV$, 
and taking $\l'_{211}=0.09$ we find using CTEQ4 \cite{SF} 
parametrization for the parton densities a cross-section of 
$\sigma(p \bar p \to \tilde \chi_1^{\pm} \mu^{\mp})=1.45pb$
at a center of mass energy $\sqrt s =2 TeV$.
Choosing other parametrizations does not change significantly   
the results since mainly intermediate Bjorken $x$ partons 
are involved in the studied process.
The cross-section depends mainly on the 
$m_{1/2}$ (or equivalently $M_2$) and $m_0$ 
soft SUSY breaking parameters.
As $M_2$ increases, the chargino mass increases
reducing the single chargino production rate. 
At high values of $m_0$, the sneutrino mass is enhanced so that 
the resonant sneutrino production is reduced. This leads to 
a decrease of the single chargino production rate 
since the $t$ and $u$ channels contributions are small 
compared to the resonant sneutrino contribution.
Finally, for values of $m_{\tilde \nu_{\mu}}$ 
(which is related to $m_0$) approaching
$m_{\tilde \chi^{\pm}_1}$ (which is related to $M_2$), a reduction of the
chargino production is caused by the decrease of the phase space 
factor associated to the decay 
$\tilde \nu_{\mu} \to \chi^{\pm}_1 \mu^{\mp}$.

The single chargino production cross-section must be
multiplied by the leptonic decays branching fractions which
are $B(\tilde \chi^{\pm}_1 \to \tilde \chi^0_1 l^{\pm}_p \nu_p)=33 \%$
(summed over the three leptons species)
and $B(\tilde \chi^0_1 \to \mu u d)=55 \%$, 
for the point chosen above of the mSUGRA parameter space.
The leptonic decay of the chargino is typically of order $30\%$ 
for $m_{\tilde l},m_{\tilde q},m_{\tilde \chi^0_2}>m_{\tilde \chi^{\pm}_1}$, 
and is smaller than the hadronic decay
$\tilde \chi^{\pm}_1 \to \tilde \chi^0_1 \bar q_p q'_p$ 
because of the color factor.   
When $\tilde \chi^0_1$ is the LSP, it decays via $\l'_{211}$ either as
$\tilde \chi^0_1 \to \mu u d$ or as $\tilde \chi^0_1 \to \nu_{\mu} d d$, 
with branching ratios $B(\tilde \chi^0_1 \to \mu u d)$ ranging between 
$\sim 40\%$ and $\sim 70\%$.


The backgrounds for the three leptons signature at Tevatron are:
(1) The top quark pair production followed by the top decays $t \to b W$
where one of the charged leptons is generated in $b$-quark decay.
(2) The $W^{\pm} Z^0$ and $Z^0 Z^0$ productions 
followed by leptonic decays of the gauge bosons.
It has been pointed out recently \cite{Match,Pai} that non negligible 
contributions can occur through virtual gauge boson, 
as for example the $W^* Z^*$ or $W \g^*$ productions. 
However, these contributions lead at most to one hard jet
in the final state in contrast with the signal and
have not been simulated.
(3) Standard Model productions as for instance the $W t \bar t$ production. 
These backgrounds have been estimated in \cite{Barb} 
to be negligible at $\sqrt s=2 TeV$. 
We have checked that the $Zb$ production gives a
negligible contribution to the 3 leptons signature.  
(4) The fake backgrounds as, 
$p \bar p \to \ Z \ + \ X, \ Drell-Yan \ +  \ X, \ b \bar b b$,
where $X$ and $b$-quarks fake a charged lepton.
Monte Carlo simulations using simplified detector
simulation cannot give a reliable estimate of this background.
(5) The \susyq background generated by the superpartner pair production.
This background is characterised by two cascade decays 
ending each with
the decay of the LSP as $\tilde \chi^0_1 \to \mu u d$ 
via the $\l'_{211}$ coupling,
and thus is suppressed compared to the signal due to
the additional branching fraction factors.
Moreover the SUSY background incurs a larger phase space suppression.
In particular its main contribution, namely
the squark and gluino pair productions, is
largely suppressed for large $\tilde{q}$ and
$\tilde{g}$ masses~\cite{tat1}.
Although a detailed estimation has not been performed we
expect that this background can be further 
reduced by analysis cuts, since at least four jets
are expected in the final state and leptons should 
appear less isolated than in the signal.


We have simulated the single chargino production 
$p \bar p \to \tilde \chi^{\pm}_1 \mu^{\mp}$ 
with a modified version of the SUSYGEN event generator \cite{SUSYGEN3}
and the \sm background
($W^{\pm} \ Z^0$, $Z^0 \ Z^0$ and $t \bar t$ productions)
with the PYTHIA event generator \cite{PYTH}. Both 
SUSYGEN and PYTHIA have been interfaced with
the SHW detector simulation package \cite{SHW},
which mimics an average of the CDF and D0 Run II detector performance.

The following cuts aimed at enhancing 
the signal-to-background ratio have been applied. First,
we have selected the events with at least three charged leptons
($e^{\pm}$ or $\mu^{\pm}$)
with energies greater than $10GeV$ for the softer of them and 
$20GeV$ for the two others, namely, 
$N_l \geq 3 \ [l=e,\mu]$ with
$E_{min}(l)>10GeV$, $E_{med}(l)>20GeV$ and $E_{max}(l)>20GeV$. 
In addition, since our final state is
$3l+2jets+ \Eslash $ we have required  
that the minimum number of jets should be equal to two, 
where the jets have an energy higher than $10 GeV$, namely, 
$N_j \geq 2$ with $E_j > 10 GeV$. 
This selection criteria suppresses the background from 
the gauge bosons production
which generates at most one hard jet. 
Note that these events requiring high energy charged leptons and jets
are easily triggered at Tevatron. 
In order to eliminate poorly isolated leptons 
originating from the decays of
hadrons (as in the $t \bar t$ production), we have
imposed the isolation cut
$\Delta R=\sqrt{\delta \phi^2+\delta \theta^2}>0.4$, where
$\phi$ is the azimuthal angle and $\theta$ the polar angle, between 
the 3 most energetic charged leptons and the 2 hardest jets. 
We have also demanded that
$\delta \phi>70^{\circ}$ between the leading 
charged lepton and the 2 hardest jets.
With the cuts described above and for an integrated luminosity 
of ${\cal L}=1 fb^{-1}$ at $\sqrt s=2 TeV$ for Tevatron Run II, 
the $Z^0 Z^0$, $W^{\pm} Z^0$, $t \bar t$ productions lead to 
$0.22,0.28,1.1$ events respectively.



\begin{figure}
\begin{center}
\leavevmode
\centerline{\psfig{figure=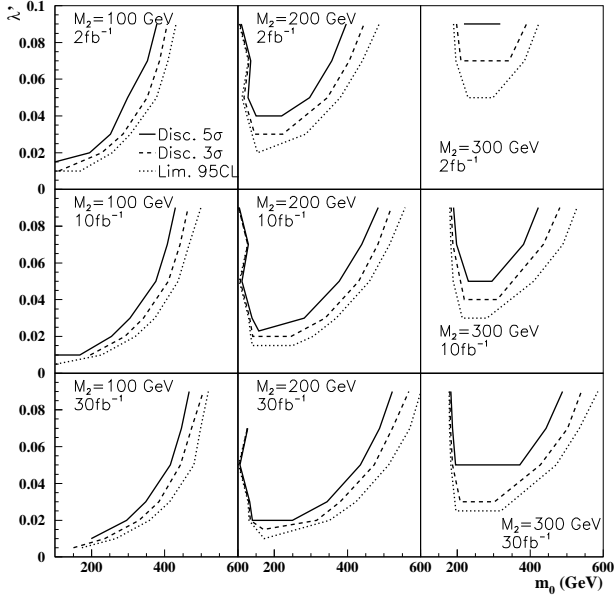,height=3.5in}}
\end{center}
\caption{Discovery contours at $5 \sigma$ (full line), $ 3 \sigma$ 
(dashed line) and limit at $95 \% \ C.L.$ (dotted line) presented
in the plane $\l'_{211}$ versus the $m_0$ parameter,
for different values of $M_2$ and of luminosity.}
\label{fig1}
\end{figure}

In Fig.\ref{fig1}, we present the $3 \sigma$ and $ 5 \sigma$ discovery  
contours and the limits at $95 \%$
confidence level in the $\l'_{211}$-$m_0$ plane,
using a set of values for $M_2$ and the luminosity. 
For a given value of $M_2$, we note that the sensitivity on the $\l'_{211}$ 
coupling decreases at high and low values of $m_0$. At high values of 
$m_0$, the sneutrino mass is enhanced 
inducing a decrease of the sneutrino production 
cross-section.
At low values of $m_0$, the sneutrino mass decreases leading to
a reduction of the phase space factor for the decay 
$\tilde \nu_{\mu} \to \tilde \chi^{\pm}_1 \mu^{\mp}$ which follows the 
resonant sneutrino production.
Similarly, we note the decrease of the sensitivity on the $\l'_{211}$ 
coupling when $M_2$ increases for a fixed value of $m_0$.
This is due to the increase of the chargino mass which  
results also in a smaller phase space factor for the sneutrino decay.

\begin{figure}
\begin{center}
\leavevmode
\centerline{\psfig{figure=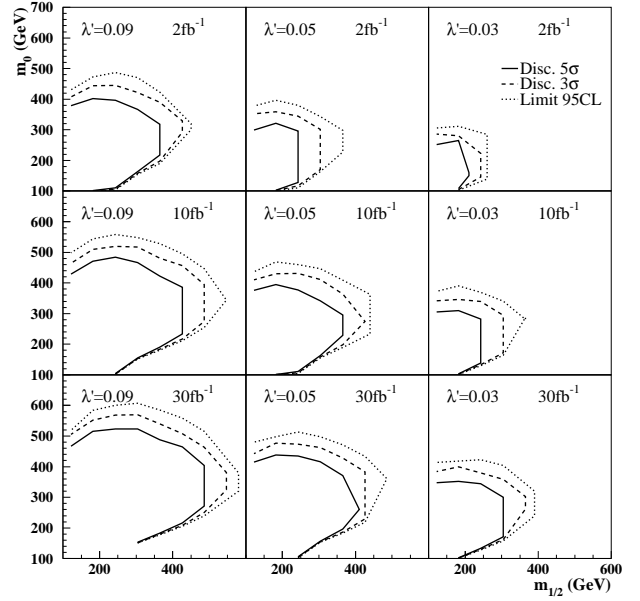,height=3.5in}}
\end{center}
\caption{Discovery contours at $5 \sigma$ (full line), 
$ 3 \sigma$ (dashed line)  
and limit at $95 \% \ C.L.$ (dotted line) presented
in the plane $m_0$ versus $m_{1/2}$,
for different values of $\l'_{211}$ and of luminosity.}
\label{fig2}
\end{figure}

In Fig.\ref{fig2}, the discovery potential is shown
in the plane $m_0$ versus $m_{1/2}$,
for different values of $\l'_{211}$ and of luminosity.
For the same reasons as above, 
we observe a reduction of the sensitivity on $\l'_{211}$ 
when $m_0$ (respectively, $m_{1/2}$ or equivalently $M_2$) 
increases for a fixed value of $m_{1/2}$ (respectively $m_0$).

\begin{figure} 
\begin{center}
\leavevmode
\centerline{\psfig{figure=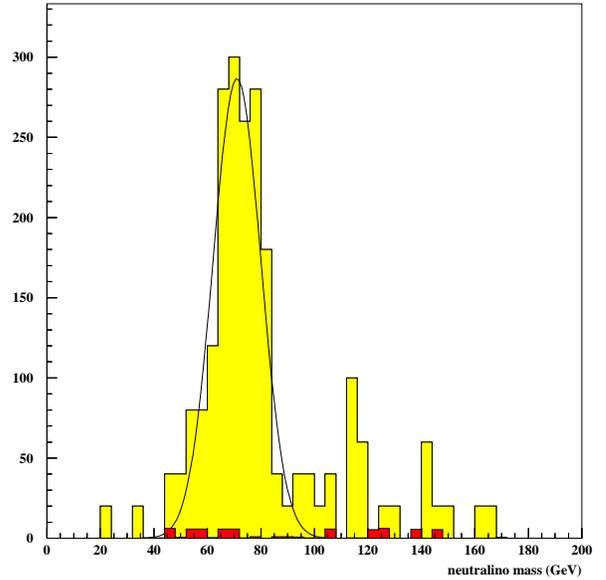,height=3.5in}}
\end{center}
\caption{Distribution for the invariant mass of
the 2 jets and the lower energy muon in the $e \mu \mu$ events, 
for a luminosity of ${\cal L}=30fb^{-1}$. The sum of the 
$WZ,ZZ$ and $t \bar t$ backgrounds is in black and the  
signal is grey. The mSUGRA point taken for this figure is, 
$m_0=200GeV$, $M_2=150GeV$ ($m_{\tilde \chi^0_1}=77GeV$), 
and the \rpv coupling is $\l'_{211}=0.09$.}
\label{fig3}
\end{figure}

An important improvement with respect to
the limits derived recently from LEP data \cite{Moriond} can 
already be obtained within the first year of Run II at Tevatron 
(${\cal L}=1 fb^{-1}$). Even
Run I data could probably lead to new limits on the \susyq parameters.
The strongest bounds on the \susyq masses obtained
at LEP in an \rpv model with non-vanishing $\l'$ Yukawa coupling are
$m_{\tilde \chi^{\pm}_1}>94 GeV$, $m_{\tilde \chi^{0}_1}>30 GeV$,
$m_{\tilde l}>81 GeV$ \cite{Moriond}. 
Note that for the minimum values of $m_0$ and $m_{1/2}$
spanned by the parameter space described in Fig.\ref{fig1} 
and Fig.\ref{fig2}, namely 
$m_0=100GeV$ and $M_2=100GeV$, the spectrum is
$m_{\tilde \chi^{\pm}_1}= 113GeV$, $m_{\tilde \chi^{0}_1}= 54GeV$,
$m_{\tilde \nu_L}= 127 GeV$, $m_{\tilde l_L}= 137 GeV$,
$m_{\tilde l_R}= 114 GeV$, so that we are well above these limits.
Since both the scalar and gaugino 
masses increase with $m_0$ and $m_{1/2}$,
the parameter space described in these figures lies outside the present 
forbidden range, in the considered framework.
\\ With the luminosity of ${\cal L}=30 fb^{-1}$ expected at the 
end of the Run II, $m_{1/2}$ values up to $550GeV$ ($350GeV$) 
corresponding to a chargino mass of about 
$m_{\tilde \chi^{\pm}_1} \approx 500 GeV \ (300 GeV)$
can be probed if the $\l'_{211}$ coupling is $0.09$ ($0.03$). 
The sensitivity on $m_0$ reaches
$600GeV$ ($400GeV$), which corresponds to a sneutrino mass of about 
$m_{\tilde \nu_{\mu}} \approx 600GeV \ (450GeV)$,
for a value of the $\l'_{211}$ coupling equal to $0.09$ ($0.03$).   
Couplings down to a value of $0.005$ can also be tested at Tevatron Run II,
in the promising scenario where $m_0 = 200 GeV$ and $M_2=100GeV$, namely,
$m_{\tilde \chi^{\pm}_1} \approx 100GeV$ 
and $m_{\tilde \nu_{\mu}} \approx 200GeV$.

Let us make a few remarks on the model dependence of our results. 
First, as we have discussed above, 
the sensitivity reaches depend on the 
SUSY parameters mainly through the \susyq mass spectrum.
Secondly, in the major part of the mSUGRA parameter space, 
the LSP is the $\tilde \chi^0_1$. Besides, in the mSUGRA model,
the mass difference between
$\tilde \chi^{\pm}_1$ and $\tilde \chi^0_1$ 
is large enough not to induce a dominant \rpv decay for the chargino.  
Notice also that we have chosen the scenario of low $\tan \beta$. 
For high $\tan \beta$, due to the slepton mixing in the third 
generation, the $\tilde \tau$ slepton mass can be reduced
down to $\sim m_{\tilde \chi^{\pm}_1}$ so that the
branching ratio of the $\chi^{\pm}_1$ decay into tau-leptons 
$\tilde \chi^{\pm}_1 \to \tilde \chi^0_1 \tau^{\pm}_p \nu_{\tau}$
increases and exceeds that into $e$ and $\mu$ leptons,
leading to a decrease of the efficiency after cuts.
For example, the efficiency at the mSUGRA point
$m_0=200GeV$, $M_2=150GeV$, $sign(\mu)=-1$, $A=0$,
is $4.93\%$ for $\tan \beta=1.5$ and $1.21\%$ for $\tan \beta=50$.
However, for still decreasing $\tilde{\tau}$ mass,
$\tilde \chi^{\pm}_1 \to \tilde \chi^0_1 \tau^{\pm}_p \nu_{\tau}$
starts to dominate over the hadronic mode
so that the efficiency loss is compensated by the leptonic
decays of the $\tau$, and the branching of the $\chi^{\pm}_1$ into
$e$ and $\mu$ leptons can even increase up to $34 \%$.
For instance, the efficiency for
$m_0=300GeV$, $M_2=300GeV$, $sign(\mu)=-1$, $A=0$,
is $5\%$ for the 2 values $\tan \beta=1.5$ and $\tan \beta=50$.


Another particularly interesting aspect of our signal is
the possibility of a $\tilde \chi^0_1$ neutralino mass reconstruction
in a model independent way. As a matter of fact,
the invariant mass distribution of 
the charged lepton and the 2 jets coming 
from the neutralino decay $\tilde \chi^0_1 \to \mu u d$ 
allows to perform a clear neutralino 
mass reconstruction. The 2 jets found in these
events are generated in the $\tilde \chi^0_1$ decay.   
In order to select the requisite charged lepton, we concentrate on the 
$e \mu \mu$ events. In those events, we know that for a relatively 
important value of the mass difference, 
$m_{\tilde \nu_{\mu}}-m_{\tilde \chi^{\pm}_1}$,
the leading muon comes from the decay, 
$\tilde \nu_{\mu} \to \tilde \chi_1^{\pm} \mu^{\mp}$,
and the other one from the neutralino decay (the electron is generated
in the decay $\tilde \chi^{\pm}_1 \to \tilde \chi^0_1 e^{\pm} \nu_e$).
In Fig.\ref{fig3}, we present the invariant mass distribution of the 
lepton and 2 jets selected through this method. 
The average reconstructed $\tilde \chi^0_1$ mass is about 
$71 \pm 9GeV$ to be compared with 
the generated mass of $\tilde \chi^0_1=77GeV$. 
In a more detailed analysis of this signal \cite{Gia,next},
the neutralino mass can be reconstructed with higher precision  
using for e.g. constrained fit algorithms.
This mass reconstruction is performed easily
in contrast with the pair production analysis in \rpv scenarios
\cite{Atlas} which suffers an higher combinatorial background. 
Moreover, a reconstruction of the chargino and sneutrino
masses is also possible.
This invariant mass distribution would also  
allow to discriminate between the signal and the SUSY background.


As a conclusion, 
we have presented a new possibility of studying resonant sneutrino 
productions in \rpv models at Tevatron. 
Results (see also \cite{next}) lead to a sensitivity on the $\l'_{211}$
coupling, on the sneutrino and chargino masses well beyond the 
present limits. Besides, a model-independent reconstruction of the  
neutralino mass can be performed easily with great accuracy.
Our work leads to the interesting conclusion that the three leptons signature
considered as a `gold plated' channel for the discovery of \susy
at hadronic colliders \cite{Match,Pai,Barb}, is also particularly 
attractive in an R-parity violation context.

We acknowledge C. Guyot, R. Peschanski and C. Savoy for 
useful discussions and reading the manuscript.

\end{document}